\documentclass[twocolumn,5p,times]{elsarticle}

\usepackage{amssymb}

\usepackage[usenames]{color}

\journal{Astroparticle Physics}

\begin{document}

\begin{frontmatter}

\title{Earth magnetic field effects on the cosmic electron flux as background for
Cherenkov Telescopes at low energies}

\author[iafe]{A.D. Supanitsky}
\ead{supanitsky@iafe.uba.ar}
\author[iafe]{A.C. Rovero}
\ead{rovero@iafe.uba.ar}
\address[iafe]{Instituto de Astronom\'ia y F\'isica del Espacio, IAFE, UBA-CONICET, Argentina}

\begin{abstract}
Cosmic ray electrons and positrons constitute an important component of the background for
imaging atmospheric Cherenkov Telescope Systems with very low energy thresholds. As the primary
energy of electrons and positrons decreases, their contribution to the background trigger rate
dominates over protons, at least in terms of differential rates against actual energies. After 
event reconstruction, this contribution might become comparable to the proton background at 
energies of the order of few GeV. It is well known that the flux of low energy charged particles 
is suppressed by the Earth's magnetic field. This effect strongly depends on the geographical 
location, the direction of incidence of the charged particle and its mass. Therefore, the 
geomagnetic field can contribute to diminish the rate of the electrons and positrons detected 
by a given array of Cherenkov Telescopes.

In this work we study the propagation of low energy primary electrons in the Earth's magnetic
field by using the backtracking technique. We use a more realistic geomagnetic field model than
the one used in previous calculations. We consider some sites relevant for new generations of
imaging atmospheric Cherenkov Telescopes. We also study in detail the case of 5@5, a proposed
low energy Cherenkov Telescope array.
\end{abstract}


\end{frontmatter}

\section{Introduction}
\label{int}

The great success achieved by current systems of Imaging Atmospheric Cherenkov Telescopes (IACTs)
has led ground-based gamma-ray astronomy to enter a period of great development, with the next
generation of instruments being designed to reach unprecedented sensitivities and angular resolution
in a more extended energy range, particularly with lower energy thresholds ($E_{th}$).
One of the main motivations for the construction of IACTs with low $E_{th}$ is the study of distant
extragalactic sources. Gamma-rays from such sources are attenuated by their interactions with the
background radiation present in the intergalactic medium. Due to this effect, the spectrum is severely
suppressed, e.g. the energy cutoff for an extragalactic object with redshift $z=1$ can be as low as
$\sim \! 50$ GeV \cite{Primack:99}. For a general discussion of the physical motivations to lower
the energy threshold of IACTs see, e.g. \cite{Albert:05}.

Current systems of IACTs like HESS, MAGIC and VERITAS have energy thresholds of $\sim \! 100$ GeV with
very large collection area. For a customized trigger system, the energy threshold of MAGIC can be as
low as $\sim \! 25$ GeV. At these energies gamma-rays can also be detected by instruments on satellites.
The Fermi LAT gamma-ray detector is able to detect photons in the energy range from $\sim \! 20$ MeV up
to more than $\sim \! 300$ GeV \cite{fermi}. Its large field of view makes it a very efficient detector
for the discovery of new sources. However, the collection area in satellites is very limited so the
sensitivity worsens very rapidly above tens of GeV. Thus, although the energy range of Fermi overlaps
with ground-based IACTs, the sensitivity at energies of few GeV is low for both detectors.

The most advanced project considering a low $E_{th}$ is the Cherenkov Telescope Array (CTA), the
largest international effort for the next generation of IACTs with an order of magnitude better
sensitivity than current systems \cite{cta}. In particular, $E_{th}$ for CTA was planned to be
$\sim \! 10$ GeV, although a more realistic value for the present design is $\sim \! 20$ GeV.
There are also specific proposals whose main technical objective is to lower the energy threshold,
like STEREO ARRAY \cite{Konopelko:05}, ECO-1000 \cite{Baixeras:04} and  $5@5$ \cite{Aharonian:01}.
For the $5@5$ array it is shown that an $E_{th}$ in the range $3-5$ GeV can be achieved by 5 IACTs
placed at 5 km of altitude. To test the feasibility of measurements at high altitudes, the OMEGA
project is also under consideration \cite{omega}.

Cosmic ray protons and electrons are the most important background sources for the discrimination
of gamma-ray showers developed in the Earth's atmosphere. Electron initiated showers are practically
indistinguishable from those initiated by gamma-rays and thus their importance for gamma-ray astronomy.
For energies below $\sim \! 20$ GeV, it is predicted that electrons could be differentiated from
gamma-ray initiated showers \cite{Sahakian:06b}.
Protons are almost two orders of magnitude more numerous than cosmic electrons at energies below
$\sim \! 10$ GeV, increasing toward higher energies as the electron spectrum is steeper than the one
corresponding to protons \cite{PAMELA:11,PAMELAprot:11}. However, protons need quite a bit more energy
to produce the same amount of Cherenkov light, specially at the lowest energies, going from a factor
$\sim \! 5$ (at $\sim \! 1$ TeV) to $\sim \! 60$ (at few GeV) \cite{Aharonian:01,Chantell:98}.
Additionally, the trigger system of IACTs discriminate against protons, although this is less effective
at the lowest energies due to fluctuations in the shower development and a drop in image intensity. In
the end, for the lowest energies, electrons become the dominant component of the differential trigger
rate for a low-energy-threshold IACT system \cite{Aharonian:01,Konopelko:05,Sahakian:06}.
All proton events passing the analysis cuts are considered as gamma-rays, and thus their reconstructed
energy will be lower than the true energy. This causes the electron dominance to diminish, making the
reconstructed flux of both electrons and protons comparable at the lowest energies \cite{Konopelko:05}.

Cosmic ray electrons and positrons\footnote{Hereafter, electrons will refer to both electrons
and positrons. Any particular case will be specified explicitly.} constitute $\sim \! 1\%$ of the
total cosmic ray flux arriving at Earth in the GeV-TeV energy range. It is believed that the high
energy component of the electron flux is directly produced by galactic sources such as supernova
remnants and pulsars \cite{Delahaye:08}. Electrons can also be produced by interactions of cosmic
ray protons or light nuclei with the interstellar medium gas.
Electrons undergo energy losses during their propagation in the interstellar medium. The main
processes are: synchrotron radiation in the galactic magnetic field, inverse Compton scattering
with photons from stars and the cosmic microwave background, bremsstrahlung with interstellar
matter, and ionization. For energies grater than $\sim \! 10$ GeV, the electron flux is dominated by the
local environment because the attenuation length is reduced to the kpc scale \cite{Delahaye:08}.

Depending on the location on Earth, the geomagnetic field can act as a shield for charged particles,
suppressing the flux of low energy electrons and, in this way, diminishing their contribution to the
background for IACTs. A first study of the geomagnetic field effects on the cosmic electron rate was
performed by Cortina and Gonz\'alez \cite{Cortina:01}, in the context of the MAGIC telescopes,
for several geographical positions around the world by using the dipolar approximation of the magnetic
field of the Earth. A better description of the geomagnetic field is provided by the International
Geomagnetic Reference Field (IGRF) \cite{IGRF:11}, which is given as a multipole expansion up to
10$th$ or 13$th$ order, depending on the version under consideration. In this case, the problem cannot
be solved analytically, and thus numerical methods are used. In particular, the backtracking technique
is used to find the allowed and forbidden trajectories \cite{Smart:00}.

The shielding of charged particles is not the only effect caused by the magnetic field of the Earth
at a given location. Extensive air showers develop in the atmosphere generating negatively and positively
charged particles, particularly electrons and positrons being the most important for IACTs. These particles
are deflected in opposite directions by the component of the geomagnetic field normal to the shower
axis. This spread causes an additional dispersion in Cherenkov images recorded by IACTs and,
consequently, diminishing the background separation efficiency. There are extensive
studies in the literature about this effect (e.g. \cite{Commichau:08}), which depends not only on
the location on Earth but also on the telescope pointing direction. While this and other effects
might be of more importance than the shielding effect, they are not considered in this paper.

In this work we study the suppression of the cosmic electron flux due to geomagnetic field effects for
locations in the Southern hemisphere. Moderate to high altitude sites with clear skies are available
only in South Africa and South America. Thus, we consider here three candidate sites in the southern
hemisphere (which are being considered for the installation of CTA): El Leoncito (31:47 S, 69:28 W),
San Antonio de los Cobres (SAC, (23:50 S, 66:16 W)), both in Argentina \cite{Rovero:08}, and the
HESS site in Namibia \cite{HESS}. We also study in detail the case of 5@5, for which we consider
the sites Llano de Chajnantor in the Atacama desert, northern Chile \cite{Aharonian:01}, and SAC
which is very close to the latter (less than $200$ km southwest).

\section{The cosmic electron flux}

The electron flux has been measured by several experiments (see \cite{PAMELA:11} for a compilation
of experimental data). The most recent measurements with better statistics are from Fermi LAT
\cite{Fermi:11} and PAMELA \cite{PAMELA:11}. Fermi LAT covers the energy range from 7 GeV to 1 TeV,
whereas the corresponding PAMELA range is from 1 GeV to 0.625 TeV. Although above $\sim \! 10$ GeV the
PAMELA spectrum seems to be softer than Fermi LAT's, they are consistent within uncertainties.
For energies below 10 GeV the data from older experiments fall within the range of the PAMELA and HEAT
\cite{HEAT} results. The latter can be taken as the extreme minimum case for the measured electron flux arriving
at Earth. For energies below $\sim \! 10$ GeV, the discrepancies on the electron flux measured by
different experiments can be explained, in part, by solar modulation effects suffered by incident electrons.

\begin{figure*}[th]
\centering
\includegraphics[width=14cm]{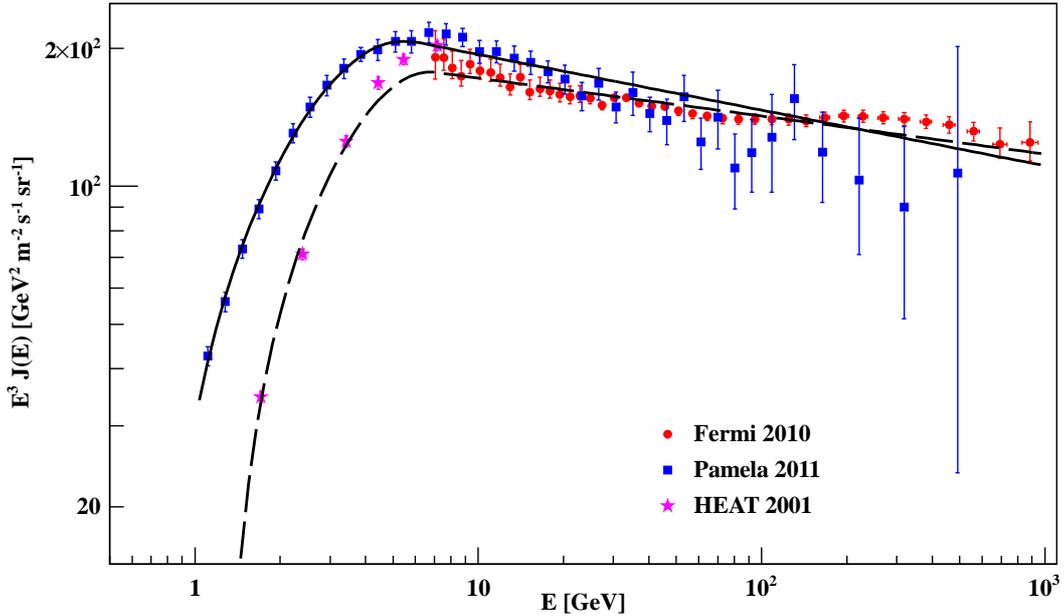}
\caption{Electron flux multiplied by $E^3$ obtained by Fermi LAT, PAMELA and the low energy part
($E\leq10$ GeV) of HEAT. Error bars include statistical and systematic uncertainties. The solid line
corresponds to the fit of data from PAMELA and Fermi LAT. The dashed line corresponds to the fit of
data from HEAT and Fermi LAT.}
\label{ElFlux}
\end{figure*}

In this work, the flux measurements from Fermi LAT, PAMELA and the low energy part ($E\leq10$ GeV) of HEAT are
considered. Figure \ref{ElFlux} shows the experimental data, where the error bars indicate statistical plus
systematic uncertainties.
For the numerical calculation an analytical expression of the electron and positron flux ($J=J_{ele}+J_{pos}$)
is used, which is obtained by fitting the data shown in figure \ref{ElFlux} with the following function,
\begin{equation}
J(E) = \left\{ 
\begin{array}{ll}
  a+\frac{b}{E}+\frac{c}{E^2}+\frac{d}{E^3}  &  E \leq 7\ \textrm{GeV} \\
                                             &                        \\ 
  \phi_0\ E^{-\gamma}                         &  E > 7\ \textrm{GeV}
\end{array}  \right.,
\label{Flux}
\end{equation}
where $E$ is in GeV, $J$ is in $m^{-2} s^{-1} sr^{-1} GeV^{-1}$, and both $\{a, b, c, d\}$
and $\{\phi_0, \gamma\}$ sets of parameters are not independent, but related by the conditions
that make the flux and its derivative continuous at $E=7$ GeV.  

Figure \ref{ElFlux} shows both fits considered here as the maximum and minimum cosmic electron fluxes:
the first corresponding to the combination of PAMELA and Fermi LAT data ($J_{PFL}$: solid line),
and the second to the combination of Fermi LAT and the low energy part of HEAT data ($J_{HFL}$:
dashed line). Table \ref{tabFlux} summarizes all parameters resulting from the fits, as specified
in equation (\ref{Flux}). It is considered that any other measured flux falls in between these
two fits so that all possibilities are covered within uncertainties. The relevance of the goodness
of these fits is discussed in next section. Nevertheless, to evaluate the geomagnetic field effects
at a given site, the fit $J_{PFL}$ is finally chosen as it is the more recent and better measured
electron flux available at present.
\begin{table*}[th]
\begin{center}
\begin{tabular}{c c c c c c c}  \hline 
Fit  & a & b & c & d & $\phi_0$ & $\gamma$  \\  \hline 
$J_{PFL}(E)$  & 1.12439 & -22.2423 & 142.483 & -90.850 & 256.183 & 3.1213   \\ 
$J_{HFL}(E)$  & 0.42764 & -12.1735 & 105.798 & -113.49 & 208.317 & 3.0829  \\  \hline
\end{tabular}
\caption{Parameters resulting from the fits of equation (\ref{Flux}) for PAMELA \& Fermi LAT
($J_{PFL}$) and for HEAT \& Fermi LAT ($J_{HFL}$).
\label{tabFlux}}
\end{center}
\end{table*}

\begin{figure}[th]
\centering
\includegraphics[width=8cm]{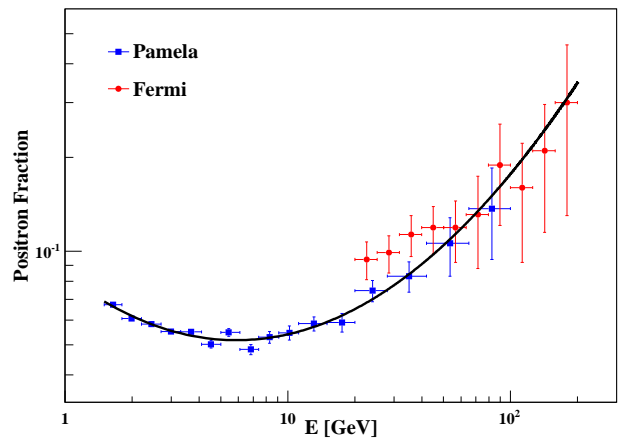}
\caption{Positron fraction as a function of primary energy measured by PAMELA and Fermi LAT. The solid line 
corresponds to a fit of the data with the function in Eq. (\ref{FitPosF}).}
\label{PosFract}
\end{figure}

Although at energies of the order of a few GeV the flux is dominated by the electron component, the 
positrons can contribute in a non negligible way to the IACTs background. Figure \ref{PosFract} shows
the positron fraction, $\delta(E) = J_{pos}(E)/(J_{pos}(E)+J_{ele}(E))$ as a function of primary 
energy, obtained by PAMELA \cite{PamelaPosF:09} and Fermi LAT \cite{FermiPosF:12}. Also shown is a 
fit of the PAMELA data with the function,  
\begin{equation}
\label{FitPosF}
\log\delta(E)=p_0 +p_1 \log E +p_2 \log^2 E,
\end{equation}
where $E$ is in GeV, $p_0=-1.078$, $p_1=-0.542$ and $p_2 = 0.352$. Note that the fit is
consistent with the Fermi LAT data and, therefore, it is reliable up to 200 GeV, the maximum energy
reached by Fermi LAT. 

\section{Numerical technique and results}

Charged particles propagating through the interstellar medium towards the Earth are deflected by 
the geomagnetic field. The deflections suffered by these particles depend on their rigidity, which 
is defined as: $R=p c/Z e$, where $c$ is the speed of light, $p$ is the momentum and $Z e$ is the 
charge of the particle. In this way, particles coming from a given direction of incidence with 
rigidities smaller than a given value are prevented to reach the Earth's surface. This minimum 
value of rigidity is called {\it rigidity cutoff} for that particular location and direction.

The dominant component of the geomagnetic field is originated in the interior of the Earth. As a 
first approximation it can be modeled by a dipole located in the center of the Earth. In this case, 
it is possible to find an analytical expression for the rigidity cutoff \cite{Stormer:30}. A better 
approximation is to consider an eccentric dipole for which it is also possible to solve the problem 
analytically. However, precise calculations require the use of more sophisticated models, like the 
one provided by the International Geomagnetic Reference Field (IGRF) \cite{IGRF:11}. In this case, 
the problem cannot be solved analytically and then numerical methods are used to obtain the trajectories 
of particles \cite{Smart:00}. In this work, the propagation of cosmic electrons in the magnetic field 
of the Earth is performed by using the program {\emph TJI95} \cite{TJI95} written by Smart and Shea. 
The trajectories are calculated by using the backtracking technique, i.e. the trajectory of an electron, 
arriving at a given geographical location, with energy $E_0$, and arrival direction $\hat{n}$, is 
calculated by propagating a positron with initial energy $E_0$ and direction $-\hat{n}$. Similarly, the
trajectories of positrons are calculated by propagating electrons. The geomagnetic field implemented
in the program corresponds to a multipole expansion up to $10th$ order (the external magnetic field,
e.g. solar, is not included) and the equations of motion are solved by using the fifth order Runge-Kutta
method. The electrons and positrons are propagated from an altitude of 20 km.

\begin{figure}[!h]
\centering
\includegraphics[width=8cm]{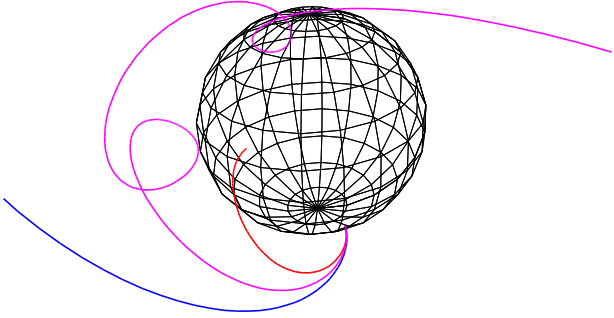}
\caption{Trajectories corresponding to electrons falling vertically at El Leoncito with energies $9$ GeV
(red line), $10.65$ GeV (magenta line), and $12$ GeV (blue line).}
\label{Traj}
\end{figure}

As an example of allowed and forbidden rigidities (or energies) for a particular location and direction,
figure \ref{Traj} shows the trajectories corresponding to electrons falling vertically at El Leoncito.
The red line corresponds to electrons with energy $E_{ele}=9$ GeV, the magenta to $10.65$ GeV, and the 
blue to $12$ GeV. As can be seen from the figure, all energies but the one corresponding to $E_{ele}=9$ 
GeV are allowed.

For an electron arriving to a given location on Earth with primary energy $E$, zenith angle $\theta$ and
azimuth angle $\phi$, the function $T_{ele}(E, \theta, \phi)$ is defined to take the value 1 if the trajectory
is allowed, and 0 if it is forbidden. Similarly, the function $T_{pos}(E, \theta, \phi)$ is defined for 
positrons.

\begin{figure}[!h]
\centering
\includegraphics[width=8cm]{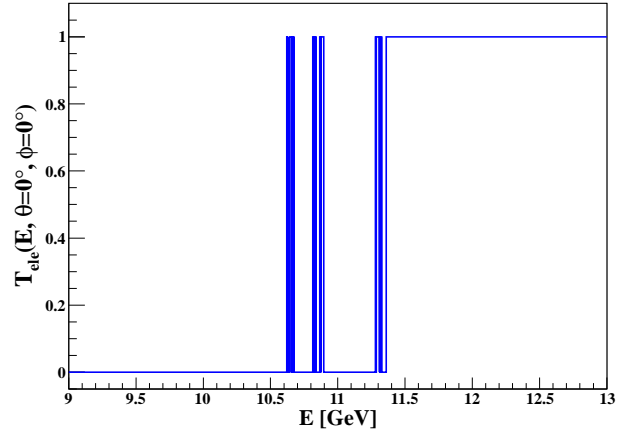}
\caption{$T_{ele}$ as a function of energy for vertical electrons falling at El Leoncito. \label{Aproba}}
\end{figure}

Figure \ref{Aproba} shows $T_{ele}$ as a function of energy for electrons, also falling vertically at 
El Leoncito. The figure clearly shows the penumbra region, which is defined as the energy transition between 
the forbidden and allowed energy regions. In the example, the penumbra is centered at about $11$ GeV and has 
a width of $\approx \! 0.8$ GeV, delimited by two energy values, $E_{L}^{ele}$ and $E_{H}^{ele}$, such that 
$T_{ele}$ is zero below $E_{L}^{ele}$ and one above $E_{H}^{ele}$. 

\begin{figure}[!h]
\centering
\includegraphics[width=8.1cm]{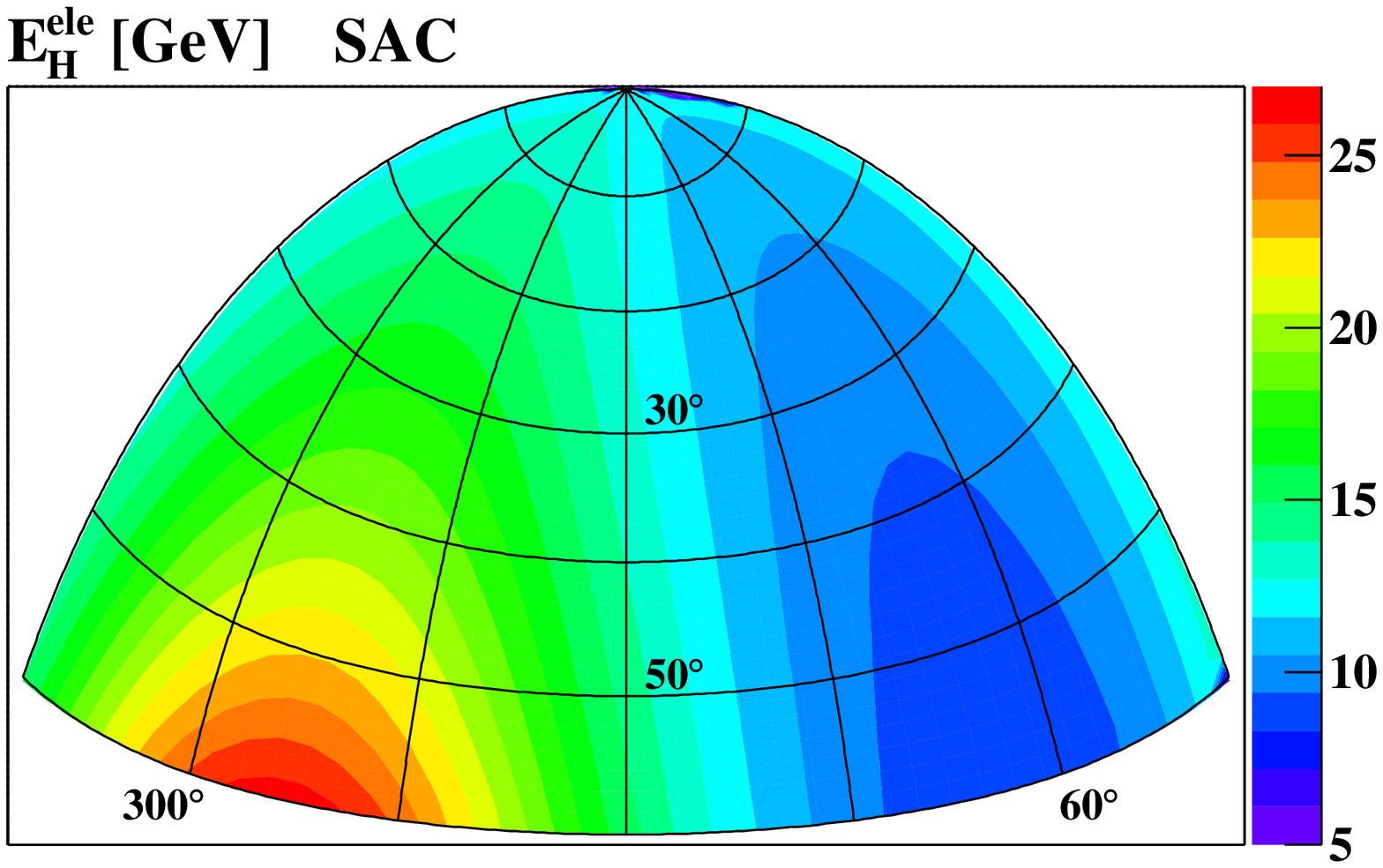}\\
\includegraphics[width=8.1cm]{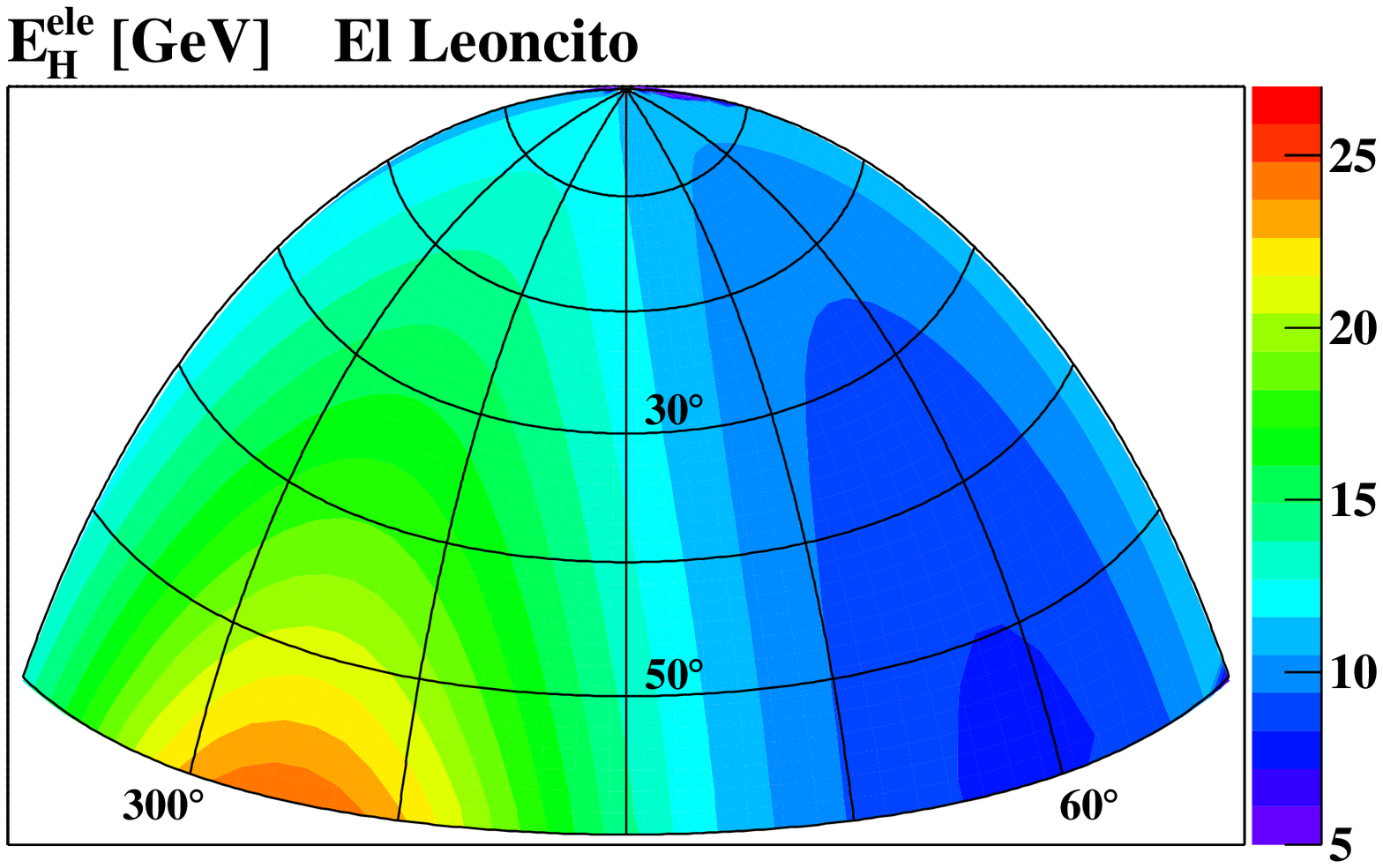}\\
\includegraphics[width=8.1cm]{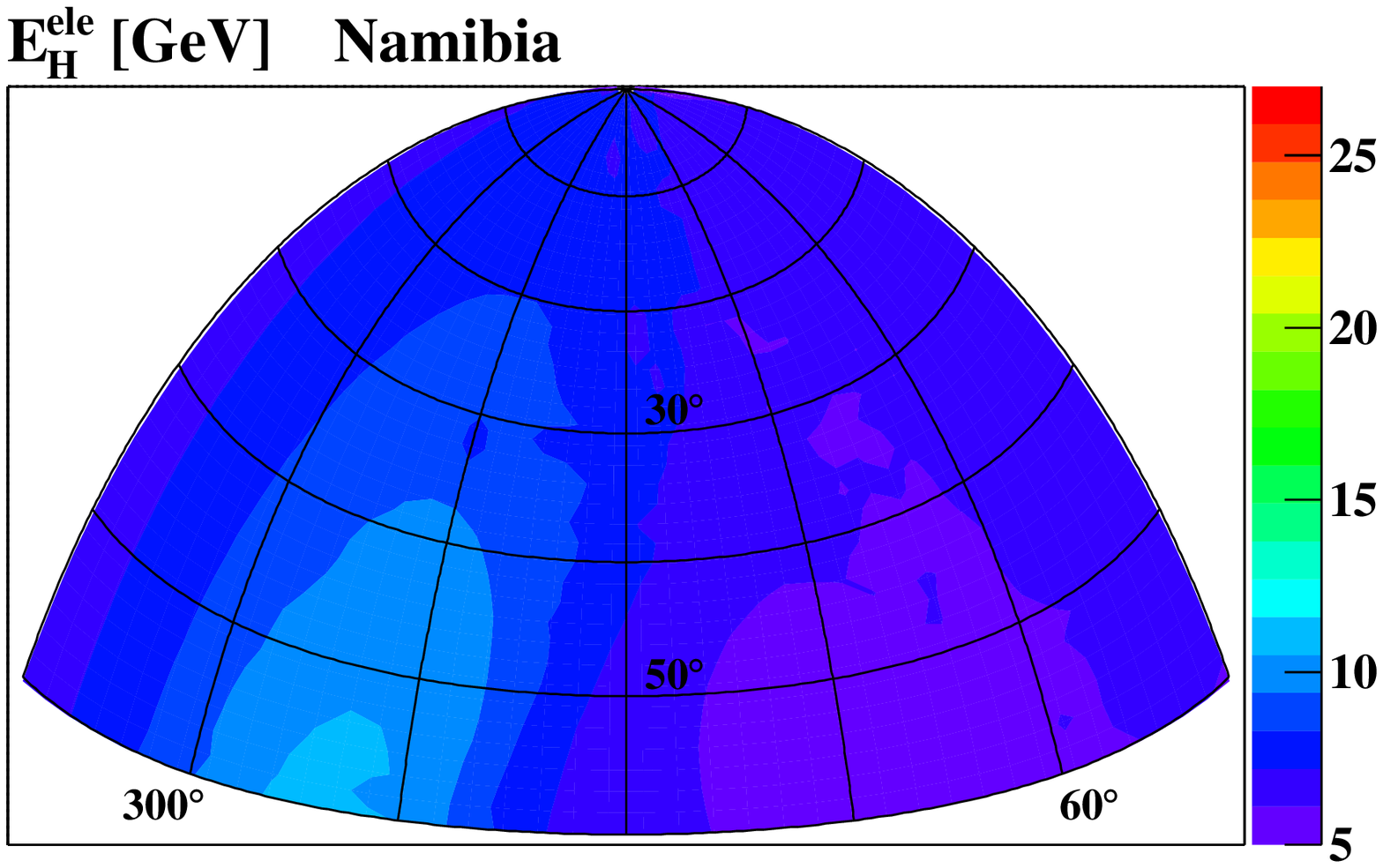}
\caption{Contour plots of $E_{H}^{ele}$ as a function of azimuth and zenith angles for the three sites under 
consideration. The azimuth angle is measured clockwise looking down from the North.}
\label{Ef}
\end{figure}

The function $T$, and also $E_{L}$ and $E_{H}$, strongly depends on the geographical location of the impact point 
and the direction of incidence. Figure \ref{Ef} shows contour plots in Aitoff projection of $E_{H}^{ele}$ as a 
function of azimuth and zenith angles, for all sites under consideration. The maximum zenith angle considered 
here is $60^\circ$. The azimuth angle is measured clockwise looking down from the North. The figure shows the well 
known East-West effect, i.e. for negatively charged particles arriving with the same zenith angle, the cutoff 
(or in this case $E_{H}^{ele}$) is smaller for the East direction. For El Leoncito $E_{H}^{ele}$ ranges from 
$\sim \! 8$ GeV to $\sim \! 25$ GeV, for SAC from $\sim \! 8.5$ GeV to $\sim \! 26.8$ GeV and for Namibia from 
$\sim \! 5.2$ GeV to $\sim \! 11$ GeV. Therefore, the largest (i.e. the best) values of $E_{H}$ correspond to SAC. 
However, the difference with El Leoncito is very small. On average, $E_{H}^{ele}$ for Namibia is about a factor  
two (or even more) smaller than the value corresponding to the other two sites.

Similar results are obtained for positrons. In this case the arrival directions with the smallest values
of $E_H^{pos}$ are in the West region, as can be seen from figure \ref{EhPosSAC} (\ref{Apos}) for SAC.

For a Cherenkov telescope with low enough energy threshold, the expected detection rate of electrons for a
given site strongly depends on the arrival direction when geomagnetic field effects are included.
In order to quantify this effect, the following parameter is introduced,
\begin{equation}
F(E_{th},\theta, \phi) = F_{ele}(E_{th},\theta, \phi)+F_{pos}(E_{th},\theta, \phi),
\label{ec2}
\end{equation}
where,
\begin{eqnarray}
F_{ele}(E_{th},\theta, \phi) &=& \frac{\int_{E_{th}}^{E_{max}} dE\ T_{ele}(E,\theta, \phi) J_{ele}(E)}{
\int_{E_{th}}^{E_{max}} dE\ J(E)}, \\
F_{pos}(E_{th},\theta, \phi) &=& \frac{\int_{E_{th}}^{E_{max}} dE\ T_{pos}(E,\theta, \phi) J_{pos}(E)}{
\int_{E_{th}}^{E_{max}} dE\ J(E)}. 
\end{eqnarray}
$F(E_{th},\theta, \phi)$ is the ratio between the detection rate of electrons with and without including 
geomagnetic field effects, for an ideal detector with detection area $A(E)=A_{0}\ \Theta(E-E_{th})$, where 
$A_{0}$ is a constant and $\Theta(x)$ is the Heaviside function (i.e. $\Theta(x)=1$ if $x\geq0$, and $\Theta(x)=0$
if $x<0$). Here $E_{max}$ is the maximum energy considered. Note that the fluxes for electrons and positrons
can be written in terms of both the total flux and the positron fraction, i.e. $J_{ele}(E)=(1-\delta(E))\ J(E)$
and $J_{pos}(E)=\delta(E)\ J(E)$. 

In order to calculate $F$, equation (\ref{Flux}) is used together with the functions $T_{ele}$ and $T_{pos}$
calculated with the {\emph TJI95} program. The maximum energy used to compute $F$ is $E_{max} = 200$ GeV,
the upper limit of the energy interval for the positron fraction reported by Fermi LAT (see Fig. \ref{PosFract}).   

\begin{figure}[!h]
\centering
\includegraphics[width=8.1cm]{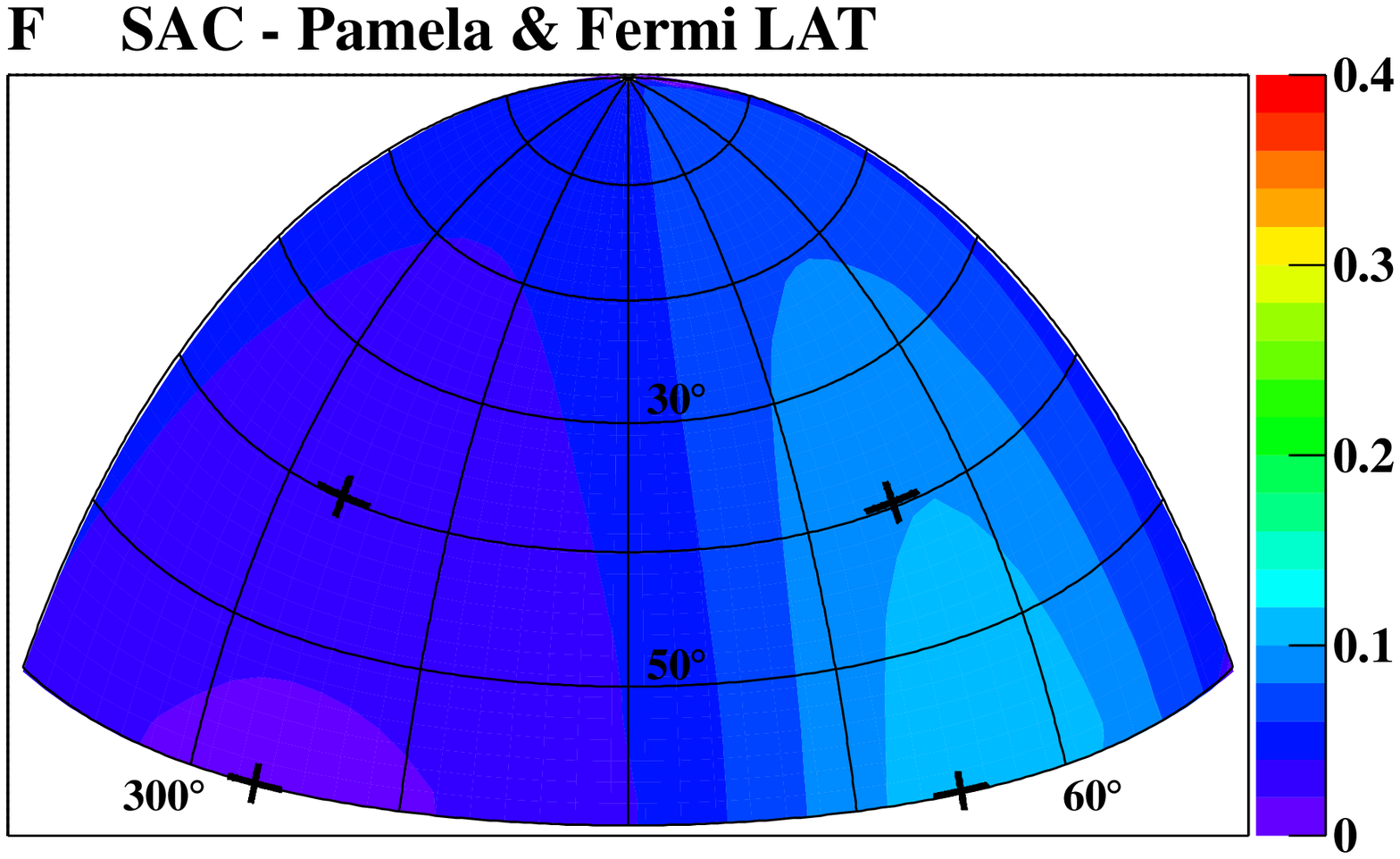}\\
\includegraphics[width=8.1cm]{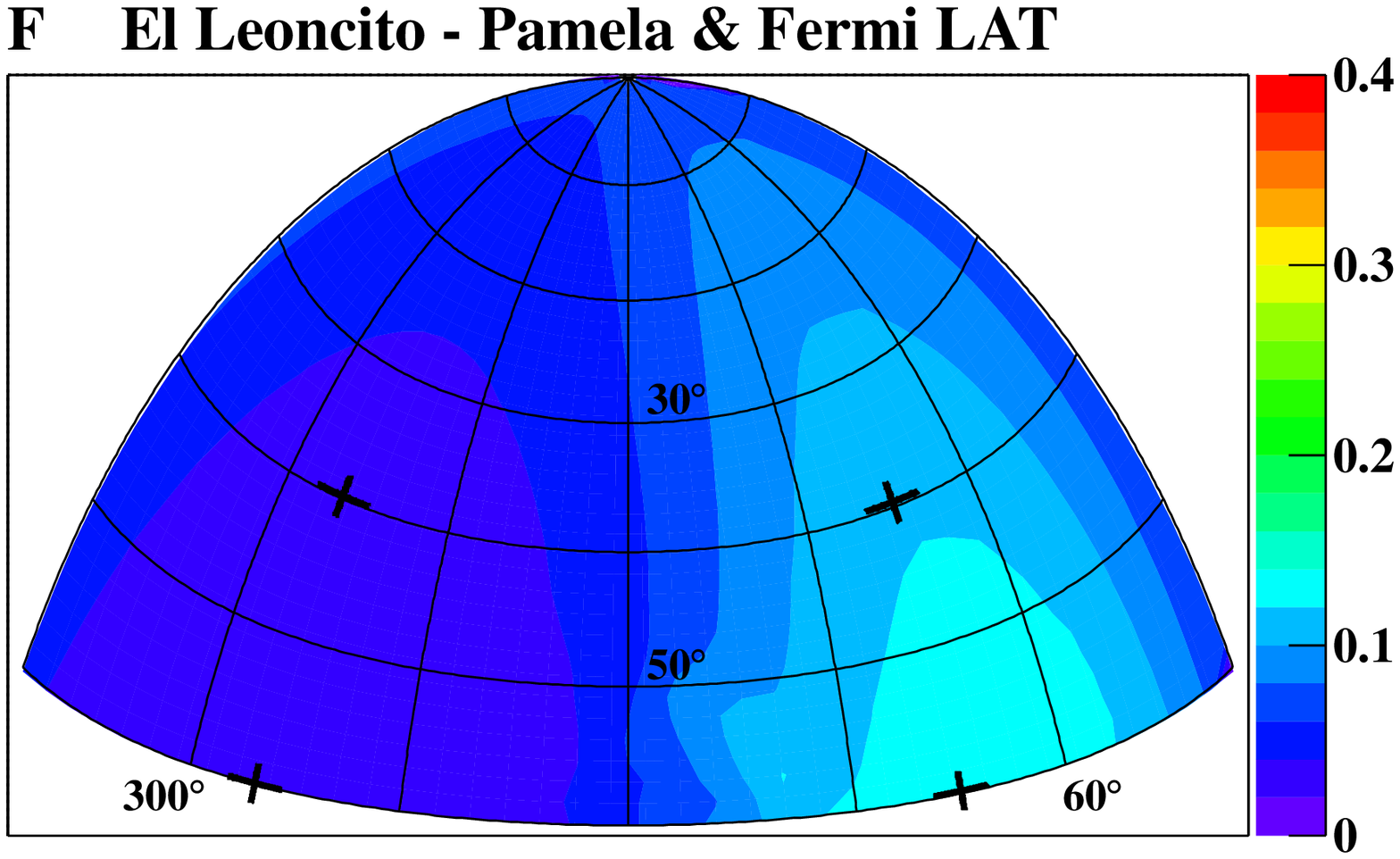}\\
\includegraphics[width=8.1cm]{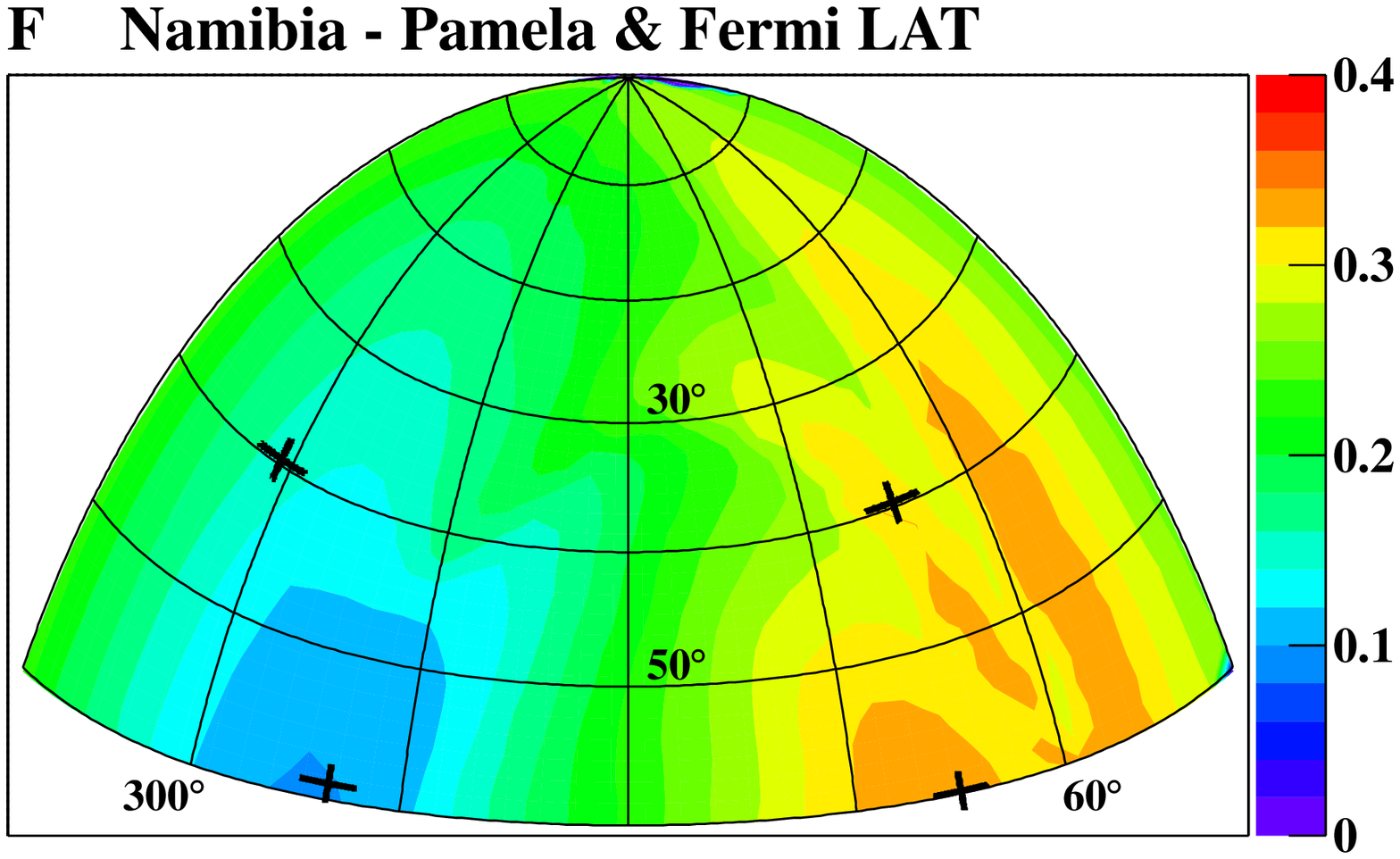}
\caption{Contour plots of $F$ as a function of azimuth and zenith angles for $E_{th}=3$ GeV. Black crosses 
indicate the directions of incidence for the maximum and minimum values of $F$, obtained for 
$\theta_{max} = 40^\circ$ and for $\theta_{max} = 60^\circ$. The electron flux used in the calculation 
corresponds to the one obtained by fitting the PAMELA and Fermi LAT data ($J_{PFL}$).}
\label{Fe}
\end{figure}

To show an example the lowest proposed energy threshold is taken, $E_{th}=3$ GeV \cite{Aharonian:01}. Figure 
\ref{Fe} shows contour plots in Aitoff projection of $F$, as a function of azimuth and zenith angles, for the 
electron flux $J_{PFL}$ (see table \ref{tabFlux}). As expected, smaller values of $F$ are obtained for 
El Leoncito and SAC. This is due to the larger values of $E_{H}^{ele}$ obtained for these two sites, in 
comparison with Namibia. Although the difference between $F$ for the first two sites is small, on average, 
$F$ is smaller for SAC. Similar results are obtained by using the fitted electron flux $J_{HFL}$ (see table 
\ref{tabFlux}).

For the characterization of a given location a study of $F$ as a function of $E_{th}$ will be presented below. 
For that study, the minimum and maximum values of $F$ are taken as representative of each location, this is 
$F_{min}$ and $F_{max}$. These quantities will correspond to two fixed positions on the sky (i.e. to given 
azimuth and zenith angles) which depend on the maximum zenith angle under consideration. The IACT technique 
is regularly used for observations at zenith angles smaller than $\sim \! 45^\circ$. For larger zenith angles 
there is a gain in collection area and a deterioration of the image quality. It is not obvious how these two 
effects change the telescope performance so simulations are needed \cite{Petry:01,Konopelko:99}. In any case, 
it is usual for IACTs to observe with zenith angles below $\sim \! 60^\circ$. Therefore, two cases are considered 
in this work in order to obtain both the maximum and minimum values of $F$ for each site: $\theta \leq 40^\circ$
and $\theta \leq 60^\circ$. These points are indicated by crosses in the cases shown in figure \ref{Fe},
for both values of $\theta_{max}$.

For $E_{th}=3$ GeV the reduction on the arriving electron rate can be significant. In table \ref{tabFe}
the minimum and maximum fractions of arriving electrons as defined in equation (\ref{ec2}) are summarized
for all locations under study, using both electron fluxes, $J_{PFL}$ and $J_{HFL}$, and for 
$\theta_{max} = 40^\circ$. As expected, all values found using $J_{HFL}$ are larger than those
obtained with $J_{PFL}$ as the latter is the largest electron flux considered, with the differences
in table \ref{tabFe} being in the range $15-24\%$. With these result, it is clear that to improve the fit
goodness for $J_{PFL}$ and $J_{HFL}$ is not relevant. Even if a much better fit is done, particularly for 
$J_{HFL}$ in the region of $E \sim \! 7$ GeV, the effect on $F$ would not be significant (estimated as 
$\lesssim \! 5\%$). Similar conclusions are reached when $\theta_{max} = 60^\circ$ is considered.
\begin{table}[h]
\begin{center}
\begin{tabular}{c c c c c c}  \hline
~ & \multicolumn{2}{c}{$F_{min}$} & & \multicolumn{2}{c}{$F_{max}$}\\ \cline{2-3} \cline{5-6}
~ & $J_{PFL}$ & $J_{HFL}$  & & $J_{PFL}$ & $J_{HFL}$ \\ \hline 
El Leoncito   & 0.031 & 0.037 & & 0.12 & 0.14 \\  
SAC           & 0.026 & 0.031 & & 0.10 & 0.12 \\  
Namibia       & 0.14 & 0.16 & & 0.33 & 0.38 \\  \hline
\end{tabular}
\caption{Fraction of arriving electrons, $F$, with energies above $3$ GeV (i.e. $E_{th}=3$ GeV). Minimum and 
maximum values of $F$ are considered for $\theta_{max} = 40^\circ$ (see text) and for both electron fluxes, 
$J_{PFL}$ and $J_{HFL}$.
\label{tabFe}}
\end{center}
\end{table}

Note that the effective area of a given array of IACTs depends on the zenith angle of the incident
gamma-ray. In order to isolate the dependence of the electron rate with the geomagnetic field,
this effect is not included in the calculations. Instead, it is discussed in section \ref{5at5}
in the context of a particular case, the 5@5 array.

\begin{figure}[!h]
\centering
\includegraphics[width=8.1cm]{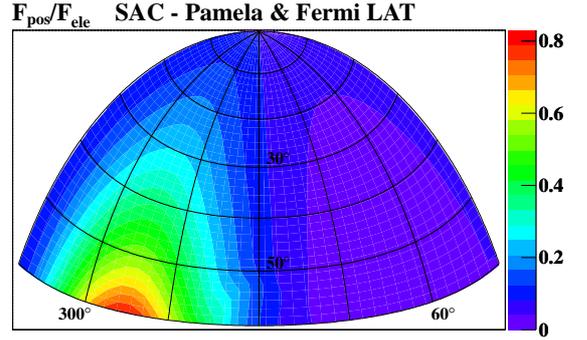}
\caption{Contour plots of $F_{pos}/F_{ele}$ as a function of azimuth and zenith angles for SAC, and using the 
electron flux $J_{PFL}$.}
\label{FPosEleSAC}
\end{figure}

As mentioned above, there are regions on the arrival direction space for which the positron component is 
important. Although the positron fraction is small at energies close to the $E_{th}$ considered here, the
positron rate can be important in regions of large values of $E_H^{ele}$ (small values of $E_H^{pos}$), 
rising the importance of the positron component. Figure \ref{FPosEleSAC} shows contour plots in Aitoff 
projection of $F_{pos}/F_{ele}$ for SAC, as a function of azimuth and zenith angles, and using the electron 
flux $J_{PFL}$. As expected, the positron contribution is more important in the West region and for large 
values of zenith angles. The maximum value of $F_{pos}/F_{ele}$ is 0.82, corresponding to the arrival 
direction $\theta \cong 60^\circ$ and $\phi \cong 270^\circ$. In the East region, where $E_H^{ele}$ 
takes the smallest values and $E_H^{pos}$ the largest ones, $F_{pos}/F_{ele}$ is smaller than $\sim 0.1$.

\begin{figure}[!h]
\centering
\includegraphics[width=8.0cm]{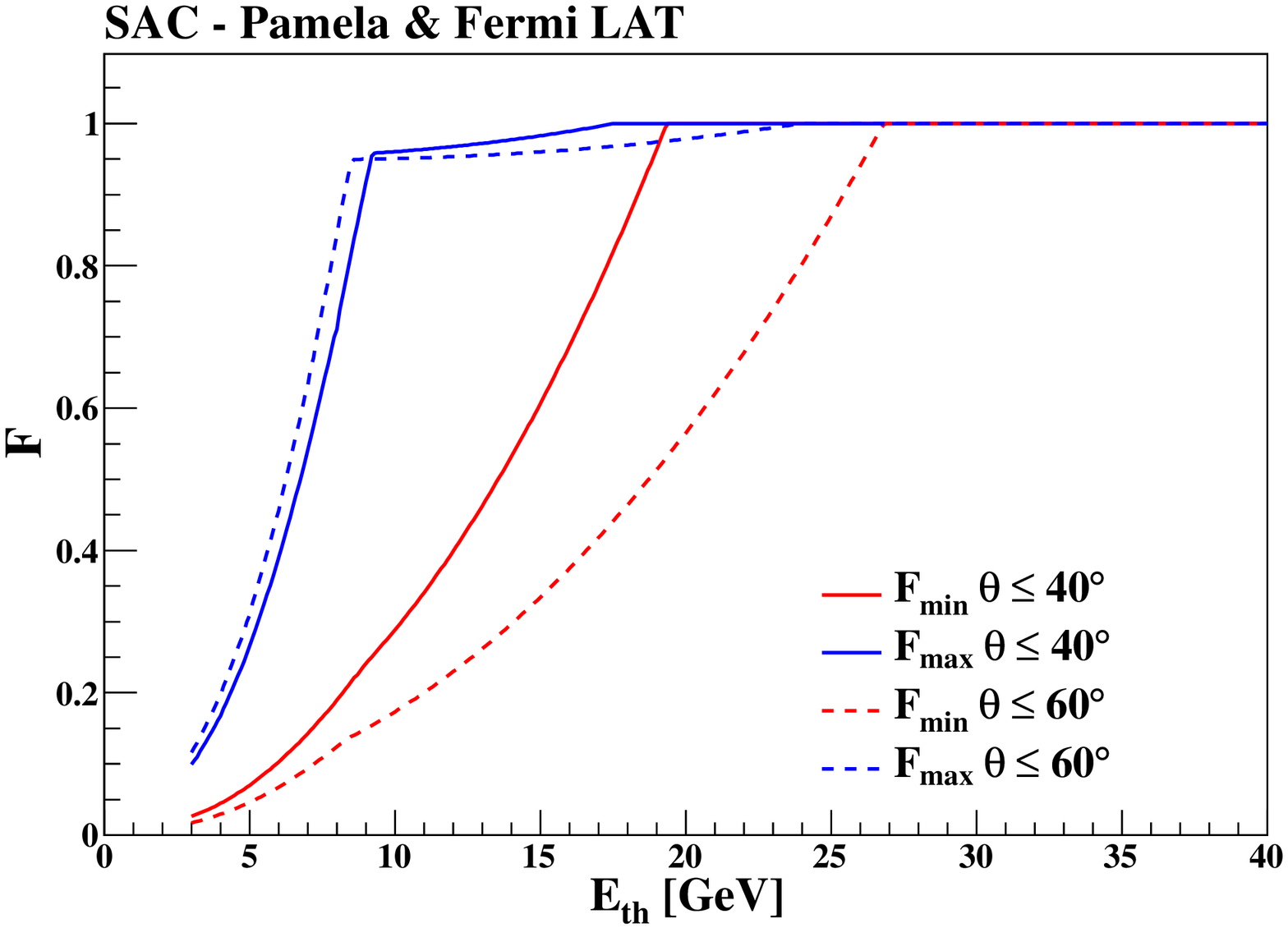}\\
\includegraphics[width=8.0cm]{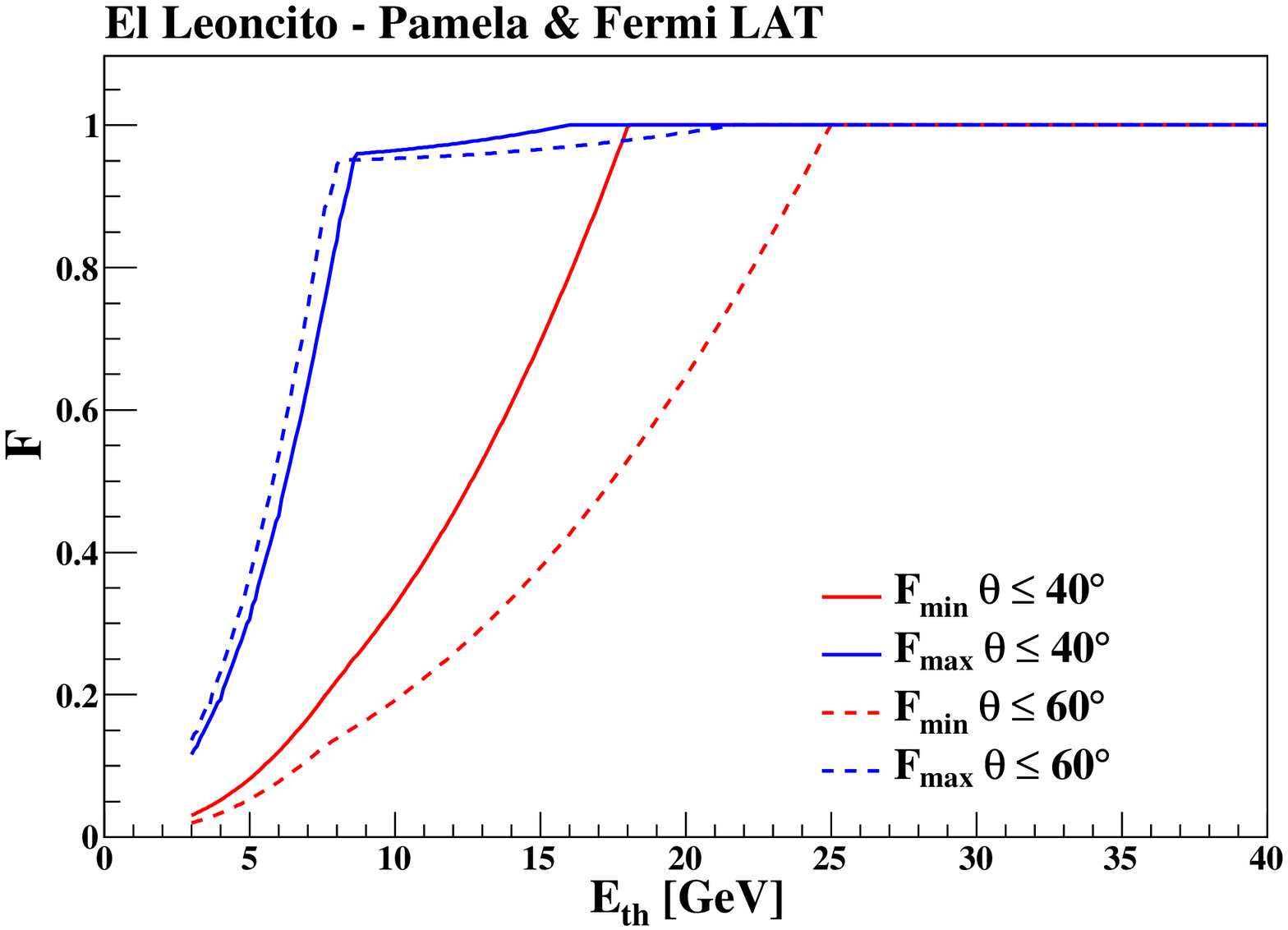}\\
\includegraphics[width=8.0cm]{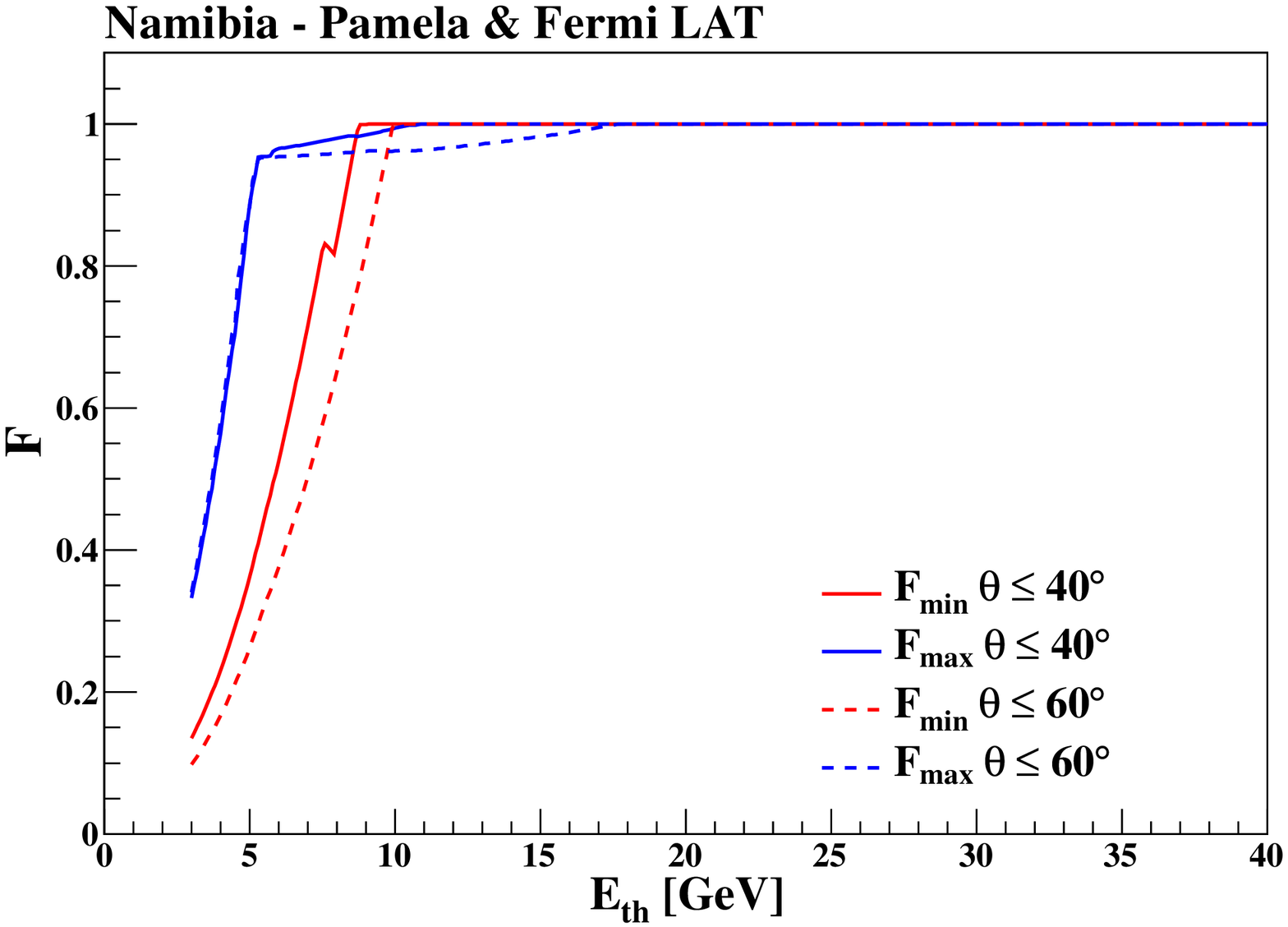}
\caption{$F$ as a function of $E_{th}$ for the three sites under consideration. Only the arrival directions 
corresponding to the maximum and minimum of $F$ obtained for $E_{th}=3$ are used (see text). The electron 
flux used corresponds to the fit of the PAMELA and Fermi LAT data. Solid lines correspond to 
$\theta_{max} = 40^\circ$ and dashed lines correspond to $\theta_{max} = 60^\circ$.}
\label{FeVsEth}
\end{figure}

The dependence of $F$ with the energy threshold is also studied. Only the arrival directions corresponding 
to the maximum and minimum of $F$ obtained above for $E_{th}=3$ GeV are used. Note that the result for any 
other arrival direction fall between these two cases. Figure \ref{FeVsEth} shows $F$ as a function of 
energy threshold for the electron flux $J_{PFL}$. Solid lines correspond to $\theta_{max} = 40^\circ$ and 
dashed lines to $\theta_{max} = 60^\circ$. From the same figure it can be seen that, if the energy threshold 
of a given Cherenkov telescope is larger than $\sim \! 27$ GeV and $\theta_{max}=60^\circ$, $F=1$ for all 
sites, i.e. the geomagnetic field do not help suppressing the electron background for any of the three 
locations. For arrays with energy threshold smaller than $\sim \! 19$ GeV ($\sim \! 27$ GeV) and 
$\theta_{max}=40^\circ$ ($\theta_{max}=60^\circ$), the best site (concerning just the suppression of the 
electron background) is SAC. Again, the results obtained for El Leoncito are quite similar to those of SAC. 
For the Namibia site, the suppression of the electron flux due to the geomagnetic field begins to be 
important for energy thresholds smaller than $\sim \! 10$ GeV, depending on the value of $\theta_{max}$.

\section{Application to the 5@5 array}
\label{5at5}

The proposed 5@5 array consists of 5 IACTs installed at an altitude of 5 km or more \cite{Aharonian:01}.
As mentioned above, the site originally proposed for the array installation is ``Llano de Chajnantor'',
Chile, at an altitude of $\gtrsim$ 5 km and $\sim$ 200 km northwest of SAC. Note that there
are several locations at 5 km of altitude very close to SAC which are suitable for the installation
of 5@5, near the location of a new astronomical facility under consideration \cite{LLAMA}.

The effective detection area of 5@5, for the zenith angle $\theta = 0^\circ$, is calculated as
\cite{Aharonian:01},
\begin{equation}
A_{eff}(E, \theta = 0^\circ) = 8.5\ \frac{\left( E/\textrm{GeV} \right)^{5.2}}{1+\left( E/5 \textrm{GeV}
\right)^{4.7}}\ \textrm{m}^2.
\label{Aeff5at5}
\end{equation}  
The effective area of any array of Cherenkov telescopes changes with the zenith angle \cite{Aharonian:08}.
The distance of the telescope to the shower maximum increases, to a good approximation, proportionally with
$1/\cos\theta$. As a consequence, the Cherenkov photons produced during the shower development illuminate
a larger area for large zenith angles. Due to this effect the effective area increases with increasing 
$\theta$ by a factor $\sim\! 1/\cos^{2}\theta$. Also, the number of Cherenkov photons that reach the telescopes
decreases with increasing $\theta$ as $\cos^{2}\theta$. This causes the energy threshold to increase by a factor 
$\sim\! 1/\cos^{2}\theta$. Therefore, the effective area as a function of energy and zenith angle can be
approximated by \cite{Aharonian:08},
\begin{equation}
A_{eff}(E, \theta) = \frac{A_{eff}(E\ \cos^2 \theta, \theta = 0^\circ)}{\cos^2\theta}.
\label{Aeff5at5theta}
\end{equation}

\begin{figure*}[th]
\centering
\includegraphics[width=8.0cm]{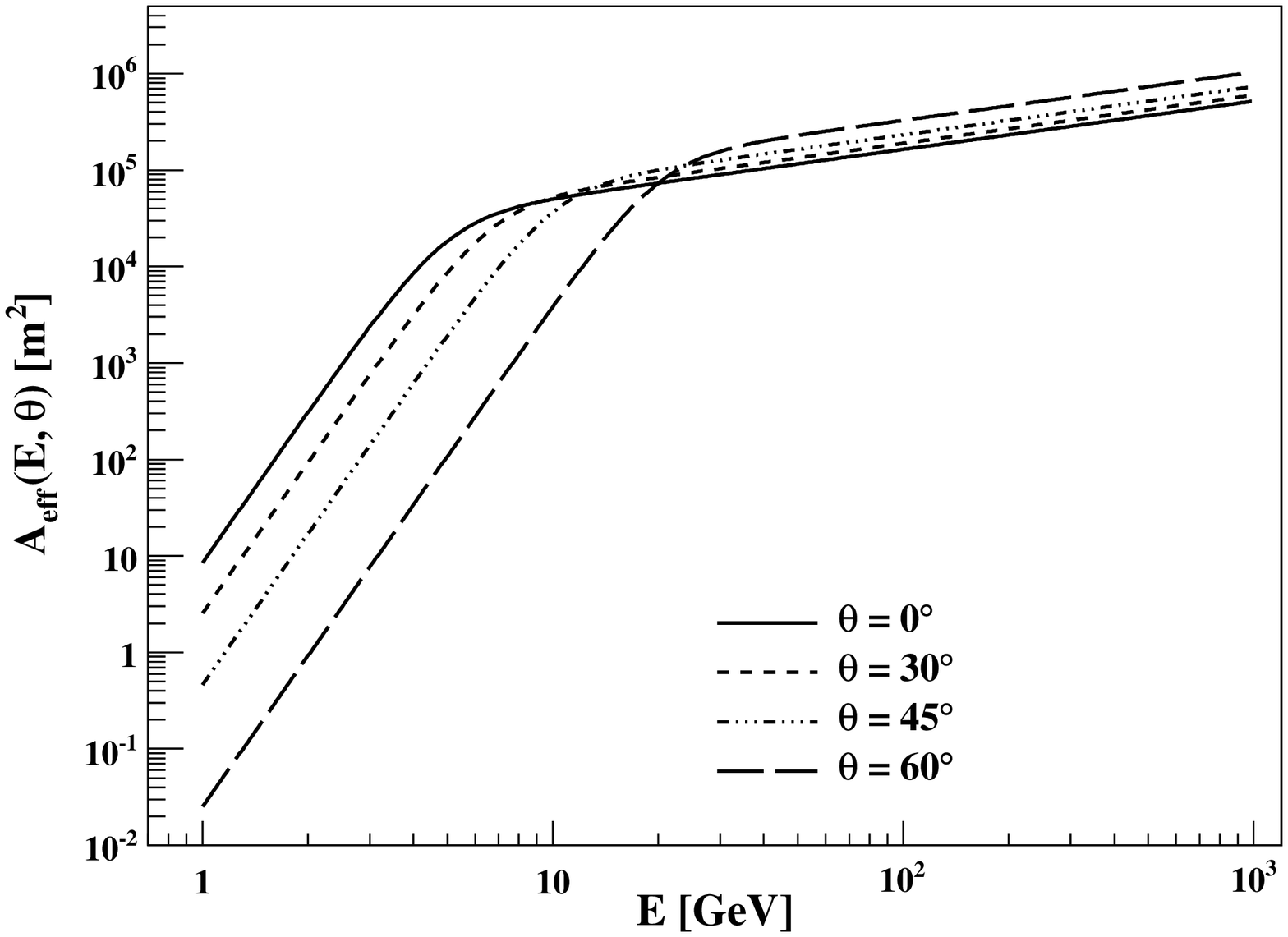}
\includegraphics[width=8.0cm]{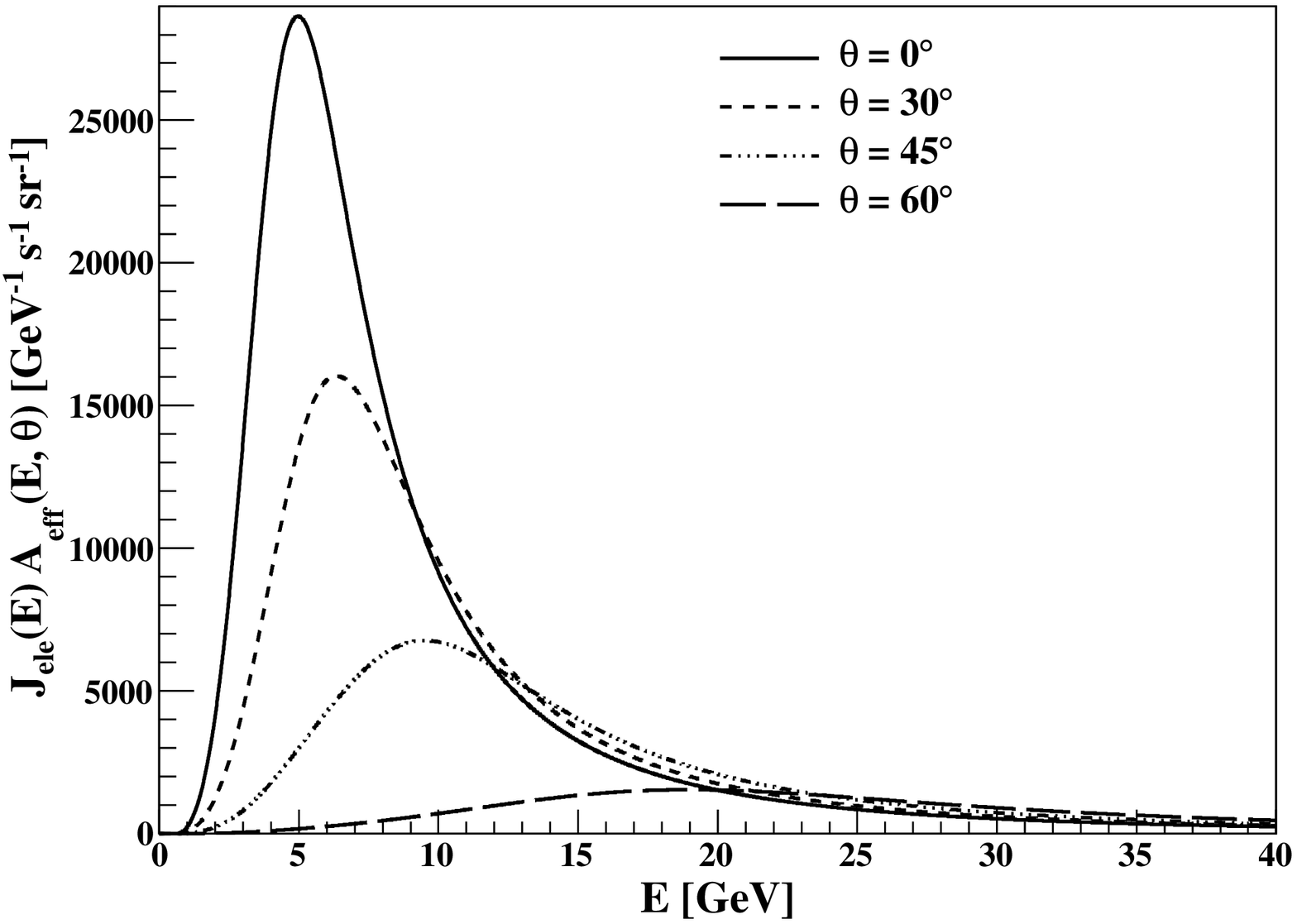}
\caption{Left panel: effective detection area as a function of energy for the 5@5 array and for
$\theta = 0^\circ, 30^\circ, 45^\circ$ and $60^\circ$. Right panel: electron flux ($J_{PFL}$)
multiplied by the effective area of 5@5 as a function of energy for $\theta = 0^\circ, 30^\circ, 45^\circ$
and $60^\circ$.}
\label{AJA}
\end{figure*}

The left panel of figure \ref{AJA} shows $A_{eff}$ for the 5@5 array as a function of energy for zenith angles
$\theta = 0^\circ, 30^\circ, 45^\circ$ and $60^\circ$, as estimated by combining equations (\ref{Aeff5at5})
and (\ref{Aeff5at5theta}). 

\begin{figure}[!h]
\centering
\includegraphics[width=8.1cm]{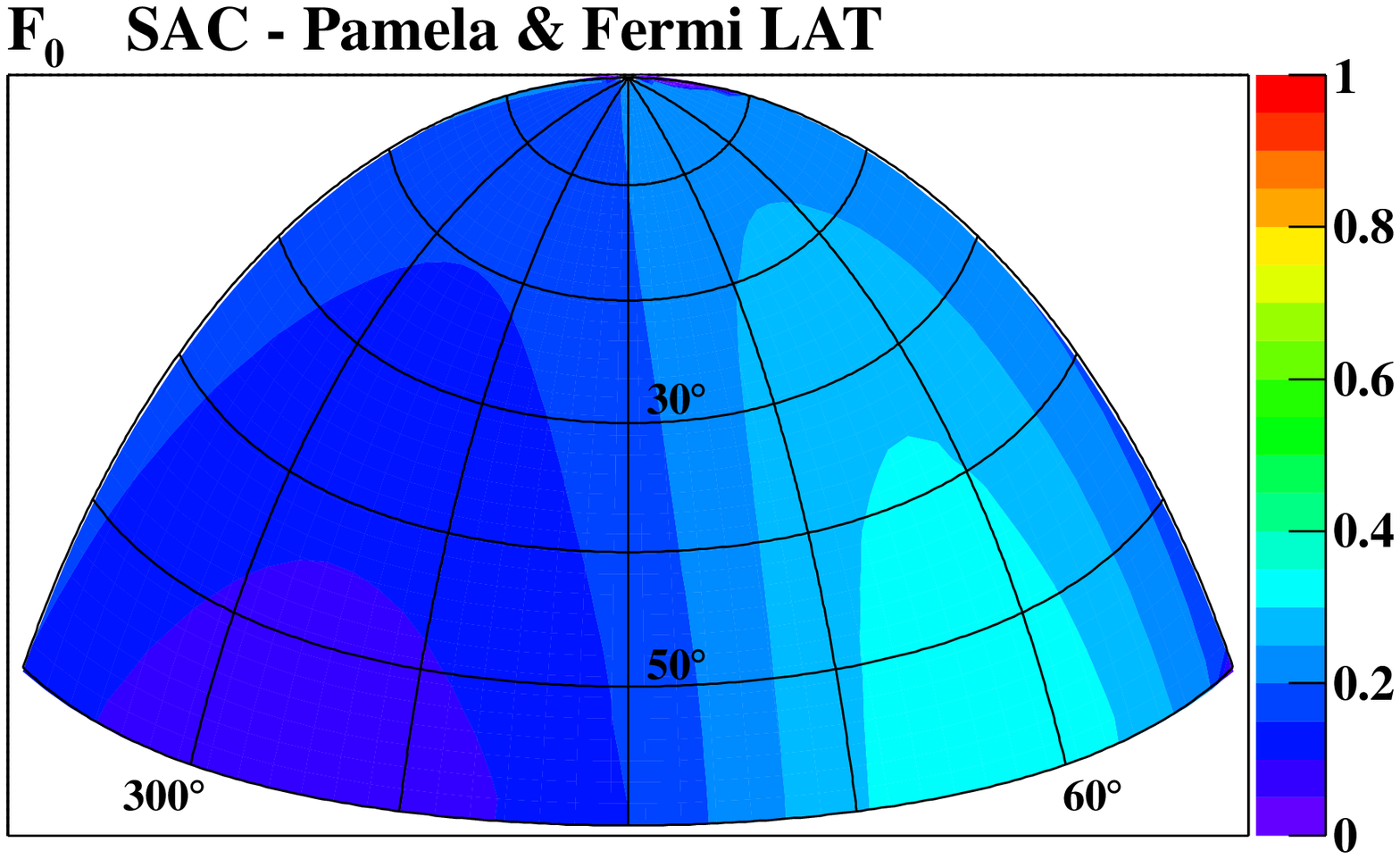}
\includegraphics[width=8.1cm]{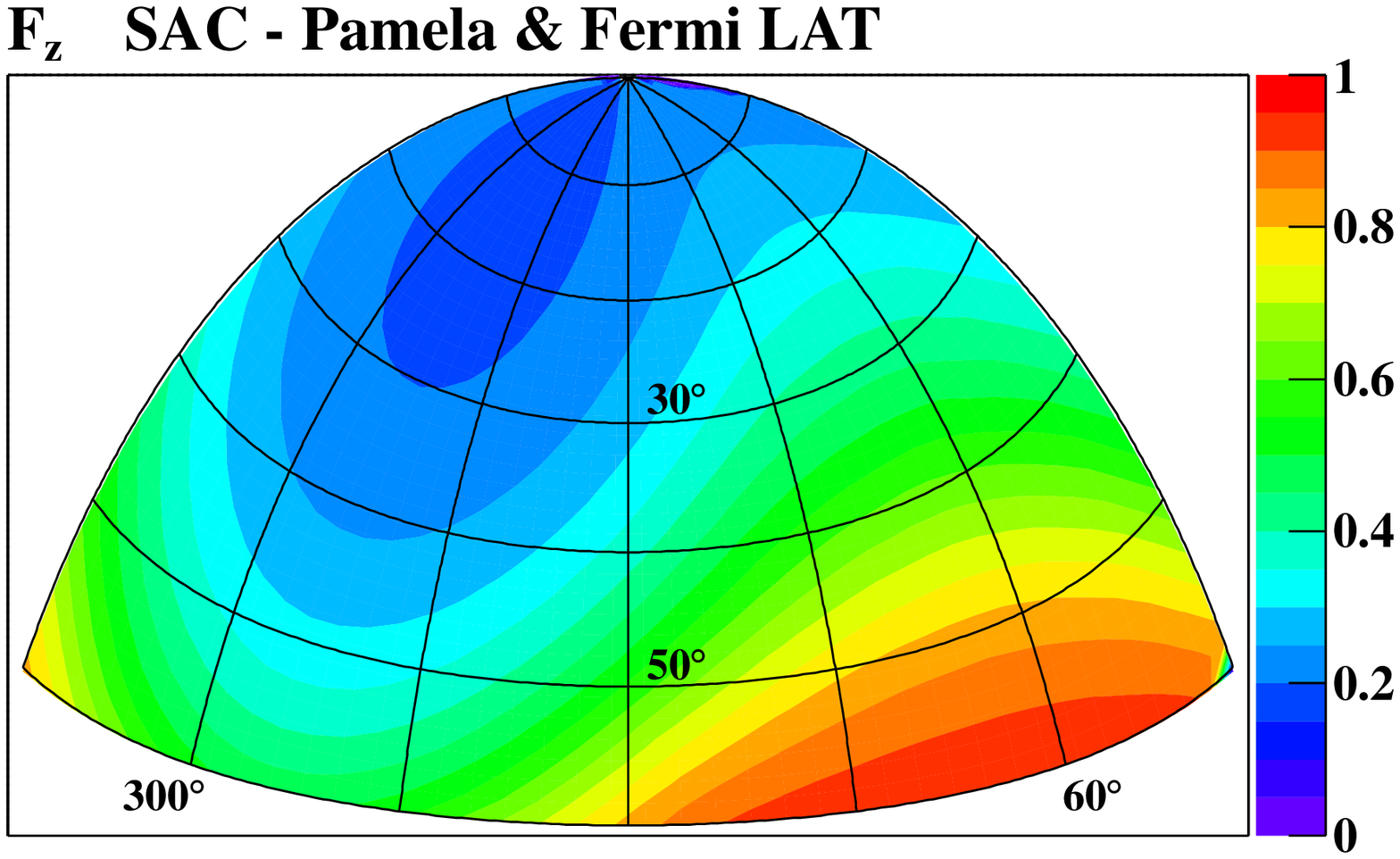}
\caption{Contour plots of $F_0$ (top panel) and $F_{z}$ (bottom panel) as a function of azimuth and zenith
angles for the 5@5 array placed in SAC, and using the electron flux $J_{PFL}$.}
\label{Fe5at5sac}
\end{figure}

In order to study the influence of the geomagnetic field on the cosmic electron rate measured by 5@5, when
it is placed in SAC, the parameter $F$ defined by equation (\ref{ec2}) is also used here. The calculation 
is performed with ($F_{z}$) and without ($F_{0}$) including the zenith angle dependence of the effective area:
\begin{eqnarray}
\label{ec5}
F_{0}(\theta, \phi) &=& F_{0}^{ele}(\theta, \phi)+F_{0}^{pos}(\theta, \phi), \\ 
\label{ec6}
F_{z}(\theta, \phi) &=& F_{z}^{ele}(\theta, \phi)+F_{z}^{pos}(\theta, \phi),
\end{eqnarray}
where,
\begin{eqnarray}
F_{0}^{\alpha}(\theta, \phi) \! \! \! \! &=& \! \! \! \! \! \frac{\int_{0}^{E_{max}} dE\ T_\alpha(E,\theta, \phi)\ J_\alpha(E)\
A_{eff}(E,\theta \! = \! 0^\circ)}{\int_{0}^{E_{max}} dE\ J(E)\ A_{eff}(E,\theta \! = \! 0^\circ)}, \\ 
F_{z}^\alpha(\theta, \phi) \! \! \! \! &=& \! \! \! \! \! \frac{\int_{0}^{E_{max}} dE\ T_\alpha(E,\theta, \phi)\ J_\alpha(E)\
A_{eff}(E,\theta)}{\int_{0}^{E_{max}} dE\ J(E)\ A_{eff}(E,\theta)}.
\end{eqnarray}
Here $\alpha=\{ele,pos\}$. Now the variables in the integrals run from zero as the collection area includes
the energy threshold. Figure \ref{Fe5at5sac} shows contour plots in Aitoff projection of $F_0$ (top panel)
and $F_{z}$ (bottom panel), as a function of arriving direction of electrons, obtained for the 5@5 array
placed in SAC and for the electron flux obtained by fitting PAMELA and Fermi LAT data ($J_{PFL}$). The values
of $F_{0}$ are larger than those obtained for the case studied in the previous section, when a step-function
effective area with energy threshold $E_{th}=3$ GeV was assumed (table \ref{tabFe}). Indeed, the maximum and
minimum values obtained in this case for 5@5 are $F_{0, min} \cong 0.11$ and $F_{0, max} \cong 0.31$, for
$\theta_{max} = 40^\circ$. This is due to the fact that for the case of 5@5 the product
$I(E,\theta=0^\circ) = J_{ele}(E)\ A_{eff}(E,\theta=0^\circ)$ has a maximum at $E\cong5$ GeV, as it can be seen
from the right panel of figure \ref{AJA}, whereas for the ideal case the maximum is reached at $E=3$ GeV.  

When the dependence of the effective area with zenith angle is included, $F_{z}$ changes even more with respect 
to the ideal case, as it is seen from the bottom panel of figure \ref{Fe5at5sac}. In fact, in the region around 
$\theta=60^\circ$ and $\phi=90^\circ$, $F_{z}$ takes values close to one. This can be understood from the behavior 
of $I(E,\theta) = J(E)\ A_{eff}(E, \theta)$, shown in figure \ref{AJA}, and $E_{H}^{ele}(\theta, \phi)$
for SAC (see top panel of figure \ref{Ef}). For $\phi=90^\circ$, $E_H^{ele}$ decreases from $\sim \! 13$ GeV
at $\theta=0^\circ$ to $\sim \! 8$ GeV at $\theta=60^\circ$, then the numerator of $F_{z}$ (equation (\ref{ec6}))
is an increasing function of $\theta$. Also from figure \ref{AJA} it can be seen that the numerator of $F_{z}$
is a decreasing function of $\theta$. As a consequence, $F_{z}$ increases with $\theta$ reaching values close
to one near $\theta = 60^\circ$. For $\phi=270^\circ$, $E_{H}^{ele}$ increases from $\sim \! 13$ GeV to
$\sim \! 25$ GeV as $\theta$ goes from $0^\circ$ to $60^\circ$, then, the numerator of $F_{z}$ decreases
with $\theta$. However, the denominator in equation (\ref{ec6}) decreases fast enough to make $F_{z}$ an
increasing function of the zenith angle.   

Similar results are obtained by using the electron flux $J_{HFL}(E)$. On the other hand, $F_{0}$ and $F_{z}$
obtained for ``Llano de Chajnantor'' are slightly smaller than the ones obtained for SAC (less than 5\%
considering all directions) because, as mentioned above, the distance between these two locations is very small
compared with the scale of variation of the geomagnetic field.  

Note that for the case of 5@5, the importance of the positron component decreases when the dependence of the
effective area with the zenith angle is included. This is due to the increase of the energy threshold of the
system with zenith angle. In particular, the maximum value reached by $F_z^{pos}/F_z^{ele}$ is $0.25$,
corresponding to the arrival direction $\theta \cong 47^\circ$ and $\phi \cong 270^\circ$.  

\section{Conclusions}

In this work we have studied the suppression of the low energy region of the cosmic electron flux,
due to the effect of the geomagnetic field, which is relevant to estimate the background for future
generations of ground-based gamma-ray detectors at the lowest energy end of the sensitivity curve.
We have studied this effect regardless its relative significance compared to other effects that might
be relevant for Cherenkov telescopes. We have estimated the cosmic electron suppression for sites that
are, or have been, proposed for the installation of new Cherenkov telescopes systems in the southern
hemisphere. 

We have considered three sites, two in Argentina (El Leoncito and San Antonio de los Cobres -- SAC) and
one in Namibia. To perform these studies we have used numerical methods based on the backtracking technique
and a multipole representation of the geomagnetic field expanded up to order 10. Considering a step function
for the collection area, we have found that the largest values of the electron energy, below which the
flux is dramatically suppressed, correspond to the site SAC. The values obtained for El Leoncito are,
on average, slightly smaller than those corresponding to SAC. We have also found that for the Namibia
site such values of energy are smaller by a factor of two, or even more, depending on the arrival direction
of the cosmic electron. We found that the total cosmic electron flux in SAC is suppressed by 90-97\% for
an energy threshold of 3 GeV, and by 0-75\% for an energy threshold of 10 GeV, depending on the arrival 
direction. For an energy threshold of 19 GeV (27 GeV) none of these locations shields the electron 
background for arrival zenith angles up to $40^\circ$ ($60^\circ$).

We have also studied in more detail the case of the proposed 5@5 array, including its simulated collection
area as a function of energy. For this case we have considered two sites: Llano de Chajnantor in the
Atacama desert, Chile, and SAC in the Argentinean Puna, where several nearby locations at 5 km of altitude
can be chosen. We have found that the geomagnetic field effect severely reduces the expected electron rate
arriving at those locations by approximately 70-90\% of the total cosmic electrons, depending on the
direction of arrival. This suppression is slightly larger for the Llano de Chajnantor site, by less than 5\%.     

\section{Acknowledgments}

The authors are members of the Carrera del Investigador Cient\'ifico of CONICET, Argentina. We thank the
anonymous referees for very helpful comments. This research was partially funded by a grant awarded
by ANPCYT, Argentina.

\appendix

\section{Upper energy cutoff for positrons}
\label{Apos}

\begin{figure}[!h]
\centering
\includegraphics[width=8.1cm]{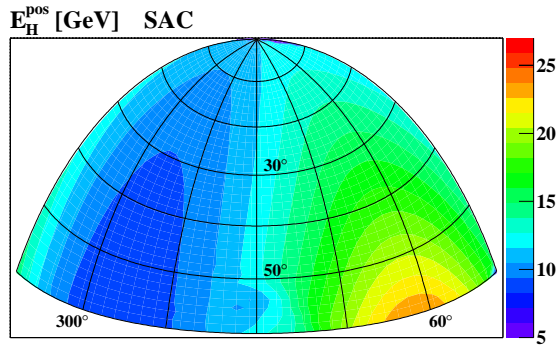}
\caption{Contour plots of $E_H^{pos}$ as a function of azimuth and zenith angles for SAC.}
\label{EhPosSAC}
\end{figure}

Figure \ref{EhPosSAC} shows contour plots in Aitoff projection of $E_{H}^{pos}$ for SAC, as a function
of azimuth and zenith angles. The smallest values of $E_{H}^{pos}$ correspond to the West region, the
opposite of what happens for $E_{H}^{ele}$.  

\end{document}